\documentclass{emulateapj}

\shorttitle{Differential Abundances in M5}
\shortauthors{Koch \& McWilliam}

\slugcomment{Accepted for publication in the Astronomical Journal}

\begin{document}

\title{A differential chemical abundance scale for the Globular Cluster M5\altaffilmark{1}}

\author{Andreas Koch\altaffilmark{2} \& Andrew McWilliam\altaffilmark{3}}
\altaffiltext{1}{This paper includes data gathered with the 6.5 m Magellan Telescopes 
located at Las Campanas Obser vatory, Chile}
\altaffiltext{2}{Department of Physics and Astronomy, University of Leicester, Leicester, UK}
\altaffiltext{3}{Observatories of the Carnegie Institution of Washington, Pasadena, CA, USA}
\email{\mbox{ak326@astro.le.ac.uk; andy@obs.carnegiescience.edu}}

\begin{abstract}
We present LTE chemical abundances 
for five red giants and one AGB star in the Galactic globular cluster (GC) M5 
based on high resolution spectroscopy using
the MIKE spectrograph on the Magellan 
6.5-m Clay telescope.  Our results are based on a line-by-line differential abundance analysis relative to the well-studied  
red giant Arcturus. The stars in our sample that overlap with existing studies in the literature 
are consistent with published values for [Fe/H] and agree to within typically 
0.04 dex for the $\alpha$-elements.  
Most deviations can be assigned to varying analysis techniques in the literature. 
This strengthens our newly established differential 
GC abundance scale and advocates future use of this method. In particular, we confirm a mean 
 [\ion{Fe}{1}/H] of $-1.33\pm0.03$\,(stat.)\,$\pm0.03 $\,(sys.) dex and also reproduce M5's 
enhancement in the  $\alpha$-elements (O,Mg,Si,Ca,Ti) at +0.4 dex, rendering M5 a
typical representative of the Galactic halo.  Over-ionization of Fe~I in the atmospheres
of these stars by non-LTE effects is found to be less than 0.07 dex.
Five of our six stars show O-Na-Al-Mg abundance patterns consistent with pollution 
by proton-capture nucleosynthesis products.
\end{abstract}


\section{Introduction}

Globular clusters (GCs) represent the oldest stellar systems in the Universe and therefore  
reflect the earliest evolutionary stages of the Galaxy. 
Although the Galactic halo GC system appears homogeneous (e.g., Cohen \& Melendez 2005; Koch et al. 2009)
and well compatible with the stellar halo in many regards (e.g., Geisler et al. 2007), 
numerous properties show broad differences between individual clusters and are at odds with halo field stars. 
These characteristics comprise anti-correlations  of the light-elements (O, Na, Al; e.g., Gratton et al. 2004) 
and the  ``second-parameter effect'', which needs to explain discordant horizontal branch (HB) morphologies 
at any given metallicity. Suggested solutions to this  problem include a broad age range in the GCs as well as
variations in their helium content (e.g., Searle \& Zinn 1978; Buonanno et al. 1997; D'Antona et al. 2002; Recio-Blanco et al. 2006). 

The Space Interferometry Mission (SIM) will take a step towards lifting such degeneracies 
by measuring accurate  parallaxes (to within $\sim$$4\,\mu$as) to numerous halo field stars and GCs. The resulting 
significant improvement in the Population~II distance scale 
will greatly reduce uncertainties in the estimated ages of the 
oldest systems in our galaxy (Shao 2004). 
An improvement of the age determinations of 
globular clusters to $\sim$5\% 
requires, however, to measure absolute metallicities of 
the clusters with an accuracy better than 0.05 dex. 
The aforementioned  problems are then resolvable through combining photometry, distance data 
and  accurate spectroscopic measurements of 
metallicities and chemical abundance ratios.

Here we present a chemical abundance study based on a 
set of high-resolution spectra of five red giants and one AGB star in the 
nearby (d$_{\sun}$=8.1\,kpc) Galactic GC M5 ($\equiv$ NGC 5904). 
Previous works have established it as a moderately metal poor system (at [Fe/H]$\sim$$-1.3$ dex) 
with heavy element abundance ratios representative of the Galactic halo (e.g., Ivans et al. 2001; Ram\'irez \& Cohen 2003; 
Kraft \& Ivans 2003; Yong et al. 2008a,b).  
While Ivans et al. (2001) suggest that the anti-correlations between those elements affected by proton-capture nucleosynthesis
in M5 are similar to those found in more metal poor GCs, Ram\'irez \& Cohen (2003) could not confirm such a trend as they did not detect very low O-abundances. 
All studies to date confirm 
a low star-to-star scatter in the iron peak- and n-capture elements of stars over a wide range of luminosities and 
thus independent of  evolutionary status. The same holds for the $\alpha$-elements with the 
exception of a broader range in [O/Fe].  

The scope of the present work is not to repeat or supplement those comprehensive measurements of chemical element ratios   
that 
already exist in abundance in the literature, but to place M5 on a highly accurate new metallicity scale. 
The importance of settling a homogeneous GC abundance scale 
is exemplified by the aberrant values for M5 based on calcium triplet 
calibrations (Rutledge et al. 1997) to the most commonly used metallicity  scales
 of Zinn \& West (1984) and Carretta \&  Gratton (1997):  for  M5, both estimates differ by (0.26$\pm$0.06) dex. 
This is far inferior to the presently achievable accuracy of high signal-to-noise (S/N),
high-resolution abundance studies in Galactic stellar populations. 
In the first paper of this series  (Koch \& McWilliam 2008; hereafter Paper~I), we  
initiated such a new GC scale by measuring chemical abundances in the GC 47~Tuc differentially to the 
well-studied Galactic K-giant Arcturus (e.g., Fulbright, McWilliam \& Rich 2006, 2007). 
By means of a line-by-line differential analysis, uncertainties in the often poorly constrained atomic parameters ($gf$-values) 
and dependencies on stellar atmospheres effectively cancel out to first order, therefore yielding absolute metallicities 
to better than 0.05 dex in [Fe/H].

In Paper~I it was shown that Arcturus suffers from an ionization non-equilibrium with
[\ion{Fe}{1}/\ion{Fe}{2}]=$-$0.08 dex; the sense and 
extent of this deviation was, however, found to differ from the red giants of similar evolutionary status and atmospheric 
parameters  in the GC 47~Tuc. In order to understand the cause for such 
discrepancies, be it systematic or a real, stellar evolutionary effect, it is crucial to understand 
whether this differential non-equilibrium is similarly present in other GCs. 

This paper is organized as follows: In \textsection 2 we present the data set and the standard reductions 
taken, while our atomic line list, stellar atmospheres are briefly introduced in \textsection 3, where we 
also recapitulate our differential abundance analysis relative to  Arcturus.
Our metallicity scale,  abundance errors and results are  
discussed in \textsection 4. Finally, \textsection 5 summarizes our findings. 

\section{Data \& Reduction}
Observations were carried out during six nights in July 2003 
using the Magellan Inamori Kyocera Echelle (MIKE) 
spectrograph at the 6.5-m Magellan2/Clay Telescope (see the observing log in Table~1). 
\begin{deluxetable*}{ccclcccccc}
\tabletypesize{\scriptsize}
\tablecaption{Log of Observations and target properties}
\tablewidth{0pt}
\tablehead{ ID & \colhead{$\alpha$ (J2000.0)} & \colhead{$\delta$ (J2000.0)} & Date & \colhead{Exposure time} & \colhead{S/N$^a$} & \colhead{V} & \colhead{$B-I$} & \colhead{$V-K_S$} & \colhead{\,v$_{r}$ [km\,s$^{-1}$]}}
\startdata
M5I-14         & 15 18 40.6 & +02 06 24 & 2003 Jul 06/23 & 3600 &  93 & 13.00 & 2.51 & 3.02 & 63.2\,$\pm$\,0.4 \\
M5I-39         & 15 18 47.2 & +02 06 20 & 2003 Jul 22	 & 2700 & 130 & 13.10 & 2.44 & 2.96 & 62.5\,$\pm$\,0.3 \\
M5III-50$^{b}$ & 15 18 24.4 & +02 01 57 & 2003 Jul 24	 & 2700 & 135 & 12.91 & 2.38 & 2.86 & 47.8\,$\pm$\,0.3 \\
M5III-94       & 15 18 27.2 & +01 59 52 & 2003 Jul 26	 & 3900 &  77 & 12.83 & 2.59 & 3.13 & 60.6\,$\pm$\,0.3 \\
M5IV-34        & 15 18 41.2 & +02 01 51 & 2003 Jul 01/06 & 2700 & 120 & 13.03 & 2.47 & 2.95 & 50.8\,$\pm$\,0.4 \\
M5IV-82        & 15 18 44.5 & +02 02 05 & 2003 Jul 06	 & 2700 & 115 & 13.23 & 2.39 & 2.93 & 48.5\,$\pm$\,0.3
\enddata
\tablenotetext{a}{S/N per pixel at blaze peak of the H$_{\alpha}$ order}
\tablenotetext{b}{AGB star (Sandquist \& Bolte 2004).}
\end{deluxetable*}
The targets for this project were pre-selected from the catalogue of Buonanno et al. (1981), from which 
we adopt the identification numbers. More recent, accurate B,V, and I-band photometry of M5 
is available from  Sandquist \& Bolte (2004) 
and complemented by infrared colors from the Two Micron All Sky Survey 
(2MASS; Skrutskie et al. 2006). These are indicated on  the color-magnitude diagram (CMD) in Fig.~1. 
\begin{figure}
\begin{center}
\includegraphics[angle=0,width=0.9\hsize]{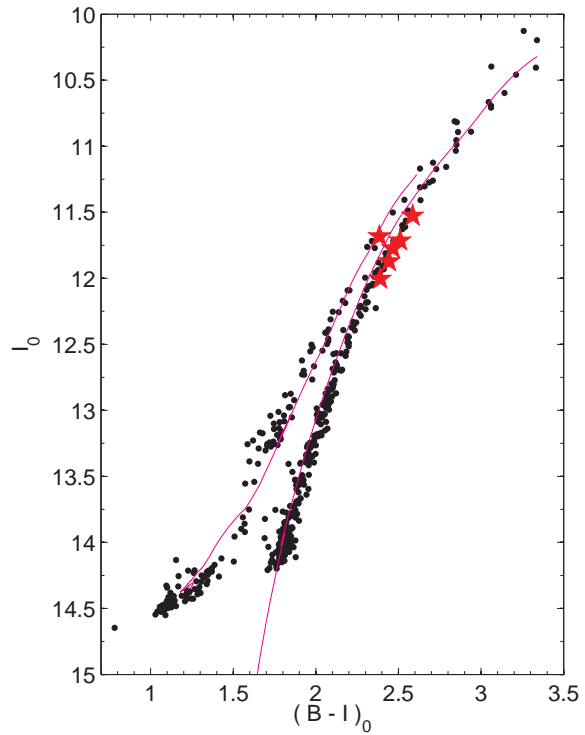}
\end{center}
\caption{Color-magnitude diagram of our targets (red star symbols) together with the RGB, AGB, and red HB data (small dots) from Sandquist \& Bolte (2004).  
 Also indicated  is a 9.7 Gyr $\alpha$-enhanced, $\eta=0.4$ Teramo isochrone with [Fe/H]=$-1.31$, shifted to E(B$-$V)=0.01 and a distance modulus of 14.55 mag. }
\end{figure}

We employed this photometry to choose giants that have 
V$-$K  photometric temperatures within $\pm$100 K of that of Arcturus, which will facilitate our differential analysis. 
Based on its location in color-magnitude space, 
one of our targets, M5III-50, was labeled as an AGB star by Sandquist \& Bolte (2004); nonetheless we retain 
it for the present work. 
None of our targets coincide with the low-resolution near-infrared calcium triplet (CaT) data set of Koch 
et al. (2006) so that no prior information on the stars'  CaT-metallicity is available. 
Four out of the six stars overlap with the high-resolution abundance analyses of Shetrone (1996), 
Ivans et al. (2001), Ram\'irez \& Cohen (2003) and Yong et al. (2008a,b).  
All these studies, however, performed element analyses using laboratory $gf$ values 
and focused on various evolutionary aspects and chemical elements so that 
it appears timely to place these stars on a homogeneous differential scale and to establish a 
firm comparison of both approaches. 
A list of the targeted stars is given in Table~1, together with their  photometric properties. 

For the present data, we utilized the full spectral coverage of the red  echelle 
set up, which yields a range of 4650--8300\AA. 
A slit with a width of 0.5$\arcsec$ and a CCD binning of 2$\times$1 pixels, 
gave a measured spectral resolving power of $R\sim$44000. 
Typically, each star was observed for 45--60 minutes, 
where we split the observations into several exposures to facilitate cosmic ray removal. 
The seeing amounted to 0.7$\arcsec$ on average, but reached as high as 1.7$\arcsec$ 
for two of the stars. 

The data reduction proceeded in exact analogy to that described in Paper I, i.e., by using the 
pipeline reduction software of Kelson et al. (2003) and subsequent continuum normalization 
through a blaze function, which was obtained 
from fitting a high-order polynomial to the high Signal-to-Noise (S/N) spectrum of a metal poor 
giant in the GC M30. 
Typically, our observing and reduction strategy results in S/N ratios of 120--130 per pixel, 
as measured in the order containing the H$\alpha$-line. Those stars affected by bad seeing conditions 
yield lower spectral quality with S/N ratios of  80--90. 

Radial velocities of the target stars were determined from the Doppler shifts of typically 35  
strong, unsaturated and unblended absorption features. Variations of the individual 
stellar velocities are, however, not critical for our analysis, since our equivalent width (EW) 
measurement program, GETJOB (McWilliam et al. 1995a), determines the line center of each feature 
independently during our later measurement process. 
Overall, we find a mean radial velocity of 54.9 km\,s$^{-1}$\ with a dispersion of 6.3 km\,s$^{-1}$, 
where all of our  targeted stars are confirmed radial velocity members (see Table~1). 
This is consistent with their proper motion based 
membership assessment by Rees (1993) and with published radial velocity data  
for those stars that overlap with our sample. 

\section{Abundance analysis}
For the present analysis we proceed in exact analogy to the methods outlined in detail 
in Paper I. Here, we briefly recapitulate the essential steps taken to apply 
the differential technique to our M5 sample.
\subsection{Line list}
We measured EWs of a large number of absorption lines using the semi-automated code GETJOB 
(McWilliam et al. 1995a). By means of a Gaussian fit to the line profiles we could achieve 
typical uncertainties of 1.8 m\AA\ ($\la$7\%), as determined from the r.m.s. scatter around 
the best-fit profile. 
Lines that appeared on adjacent spectral orders 
could usually be well measured to within this uncertainty and as the final value we adopted the error-weighted mean. Line-free continuum regions were taken from the study of 
Fulbright et al. (2006), which are, in turn, based on a detailed measurement of the metal rich giant 
$\mu$~Leo. 
The line list for the present work is identical to that in Paper I and was taken from Fulbright et al. 
(2006) for the iron lines and Fulbright et al. (2007) for the $\alpha$-element transitions.
These authors had optimized this list to exclude any contamination from blended features. 
Here we employ the same line list for consistency with our earlier work.

Previous measurements (e.g., Carretta \& Gratton 1997; Ivans et al. 2001) indicate that M5 is 
more metal poor by ca. 0.5 dex on average than 47 Tuc  (Koch \& McWilliam 2008) and 0.7 dex more metal poor than our reference star, Arcturus (Fulbright et al. 2006).  
Thus, there may be fewer strong absorption lines available for measurement in our M5 stars. 
However, we found $\sim$75 \ion{Fe}{1} lines from our list with sufficient strength for
reliable EW measurement and accurate abundance determination. 
Also the $\alpha$-elements' absorption features from the line list of Paper~I 
show up clearly in the spectra and there was no need to supplement the line list from other sources. 
Thanks to the lower metallicity, we could have included extra lines that are clean and unsaturated
in our M5 stars, but blended or saturated in Arcturus. However, for such lines we could not 
perform line-by-line differential abundance analysis relative to Arcturus.
The EWs of our lines 
for both Arcturus and the Sun, which are essential to place our GC measurements on the 
solar abundance scale, are also taken from the published measurements of Prochaska et al. (2000)  
and Fulbright et al. (2006, 2007). 
Hyperfine structure splitting for Na and Al was not included in our analyses, since the 
corrections are negligible in our red giant spectra. 
The final list with our measured EWs is shown in Table~2. 
\begin{deluxetable*}{cccrrrrrr}
\tabletypesize{\scriptsize}
\tablecaption{Line List}
\tablewidth{0pt}
\tablehead{ \colhead{} & \colhead{$\lambda$} &  \colhead{E.P.} &  \multicolumn{6}{c}{Equivalent Widths [m\AA]} \\
\cline{4-9}
 \raisebox{1.5ex}[-1.5ex]{Ion} & \colhead{[\AA]} &  \colhead{[eV]} & \colhead{M5I-14} & 
\colhead{M5I-39}  & \colhead{M5III-50} & \colhead{M5III-94} & \colhead{M5IV-34} & \colhead{M5IV-82}  }
\startdata
Fe I  & 5432.95  & 4.45 & 68.0 & 57.4 &  \nodata & 62.7 & 53.6 & 55.1 \\
Fe I  & 5460.87  & 3.07 & 10.0 & 16.0 & 13.6  & \nodata  & 11.0 & 10.6 \\
Fe I  & 5462.96  & 4.47 & 82.8 & 80.8 &  \nodata  &103.0  &84.3 & 86.0 \\
Fe I  & 5466.99  & 3.57 & 38.1 & 37.0 & 22.9  &47.6 & 34.6 & 35.3 \\
Fe I  & 5470.09  & 4.45 & 16.1 & 13.5 & 13.9 & 22.1 & 15.3 & 15.6 
\enddata
\tablecomments{This Table is published in its entirety in the electronic edition of the {\it Astronomical 
Journal}. A portion is shown here for guidance regarding its form and content.}
\end{deluxetable*}
\subsection{Stellar atmospheres and parameters}
Stellar abundances for each of the absorption lines were computed using the {\em abfind} 
driver of the 2002 version of the synthesis program MOOG (Sneden 1973). 
The stellar atmospheres for this analysis were generated from the 
Kurucz LTE models\footnote{\url{http://kurucz.harvard.edu}} without convective overshoot. Both Arcturus and M5  are enhanced in the $\alpha$-elements by approximately $+0.4$\,dex 
(Arcturus; Peterson et al. 1993; Fulbright et al. 2007) and $+$0.3 dex (M5; Ivans et al. 2001; 
Ram\'irez \& Cohen 2003; Yong et al. 2008a,b), 
 which prompted us to use the  $\alpha$-enhanced opacity distributions 
AODFNEW\footnote{\url{http://wwwuser.oat.ts.astro.it/castelli}} by F. Castelli.

For Arcturus we employed the same EWs and abundances for  individual lines as in Paper I. 
These were based on its well-established parameters (T$_{\rm eff}$=4290 K, log $g$=1.64, $\xi$=1.54 
km\,s$^{-1}$, [$M$/H]=$-$0.49 dex). 
Note, however, that a persistent problem with ionization equilibrium in Arcturus was found in Paper~I, which  will affect a proper discussion of the respective equilibrium in the M5 stars.  We will 
return to this aspect in detail in Sect.~3.3.  

Effective temperatures ($T_{\rm eff}$) for the M5 stars  were calibrated  using the empirical  V$-$K color-temperature  
relation of Alonso et al. (1999). To this end, we exploited the V-band photometry of 
Sandquist \& Bolte (2004), complemented by infrared K-band magnitudes from 2MASS (Skrutskie et al. 2006).  
In order to convert the photometric sets into the TCS filter system, needed for the Alonso et al. (1999) calibrations, 
we applied the proper transformations from Alonso et al. (1998) and the Explanatory Supplement to the
2MASS All Sky Data Release (Cutri 2003).   

For our initial reddening value we adopt 0.03$\pm$0.01 mag, which is fully consistent with 
the values provided by the Schlegel et al. (1998) maps and adopted by Catelan (2000) and 
Recio-Blanco et al. (2005) 
from the HB. 
In addition, we use the extinction law of 
Winkler (1997) throughout this work.
As a result, uncertainties on the individual T$_{\rm eff}$ from photometric 
and calibration errors, are of the order of 30 K.
As noted in Paper I, in our differential analysis it is necessary to add a zero point shift of +38 K 
to the M5 giant temperatures, due to a difference in the Alonso et al. (1999) color calibration 
for Arcturus compared to its value from angular diameter measurements. 

In addition, we calibrated spectroscopic temperatures by eliminating any trend in the 
plot of (differential) \ion{Fe}{1} abundance vs. excitation potential (to wit, excitation plot; 
see Fig.~2, top panel). 
Since each abundance datum in M5 is truly differential to the same absorption line in 
Arcturus, our spectroscopic temperatures are on the same physical T$_{\rm eff}$ scale 
as our reference star. Only lines with EW$>$20 m\AA\ were utilized in this iteration step. 
On average, the difference between color- and excitation temperatures, T(V$-$K)$-$T(spec), 
is  $-$2 K with an 
r.m.s. scatter of 20 K so that the random error on the mean difference from the six targets 
is a mere 8.3 K. In this context, a reddening change of 0.01 in E(B$-$V) corresponds to 
a shift in temperature of $\sim$19 K, so that the scatter in the difference between 
photometric and spectroscopic 
T$_{\rm eff}$ values can well be accounted for  by the reddening uncertainties cited in the literature. 

In practice, we used the average of the spectroscopic and photometric T$_{\rm eff}$ estimates
in the subsequent abundance analyses. 
It is reassuring that the temperatures  of those four stars in common with recent high-resolution studies 
agree well within the respective uncertainties -- our values are cooler 
by $-$7 K on average, with an r.m.s. scatter of 28 K. 

Surface gravities, log\,$g$, for the M5 stars are derived from the canonical stellar structure equations 
(Eq.~1 in Paper~I) and assume the  previously determined T$_{\rm eff}$ and  standard Solar parameters.  
For the luminosities we used the dereddened V-band photometry of Sandquist \& Bolte (2004) 
with bolometric corrections from the Kurucz website and a distance modulus to M5 of 
(m$-$M)$_0$=14.55$\pm$0.10 mag. The latter value is an average from the studies of   
Reid (1997); Gratton et al. (1997); Layden  et al. (2005) and Recio-Blanco et al. (2005) and based on various methods 
such as Hipparcos parallaxes, the white dwarf sequence, and fits of the main sequence turn-off or a zero-age horizontal branch.

Stellar masses of the 
target M5 stars are based on a comparison with the most recent  Teramo 
$\alpha$-enhanced isochrones (Pietrinferni et al. 2004) 
with $\alpha$ element opacities from Ferguson et al. (2005). 
This is illustrated in Fig.~1, where we compare the observed CMD with the $\alpha$-enhanced, 
$\eta=0.4$ 
Teramo isochrone with age and metallicity of 9.7 Gyr and $-$1.31 dex, respectively. 
These values provided a reasonable representation of M5's RGB, AGB and its red HB. 
For ages in the range of 8--10 Gyr, as found in the literature (De Angeli et al. 2005; Meissner \& Weiss 
2006), the isochrones were interpolated to yield stellar masses 
between 0.83 and 0.88 M$_{\sun}$. 
In practice, we adopt as RGB mass the value from the 9.7 Gyr track, viz. 
(0.84$\pm$0.05 M$_{\sun}$), and 0.69 M$_{\sun}$ for the AGB star.
The quoted error on mass accounts for uncertainties in 
distance modulus, reddening, V-band photometry, BC and the permitted  age range. 

Accordingly, the log $g$ values given in 
Table~3  could be determined to within $\pm$0.06 dex, based on individual uncertainties 
in each of the contributions discussed above, where the main factor is the 
distance modulus uncertainty of 0.10 mag.    
Moreover, the four stars in common 
with other the studies agree well with our parameters (at a mean deviation of 0.05 dex and r.m.s. 
scatter of 0.02 dex)\footnote{Note, however,  that Ivans et al. (2001) find differences 
between their photometric and spectroscopic gravities of 0.26 dex on average.}.

The microturbulent velocity $\xi$ was initially set to that of Arcturus (1.54 km\,s$^{-1}$), 
and subsequently iterated by demanding a zero slope in the plot of differential abundance 
with EW (bottom panel of Fig.~2).  A linear fit to the data then fixed $\xi$ to within 0.05 km\,s$^{-1}$. Although the final 
abundances are fairly insensitive to the actual value for $\xi$ (see Sect.~5), it is comforting that 
all our microturbulent values agree with those found in the literature, where the mean deviation is 
0.02  km\,s$^{-1}$ with an r.m.s. scatter of  0.10 km\,s$^{-1}$.
\begin{figure}
\begin{center}
\includegraphics[angle=0,width=1\hsize]{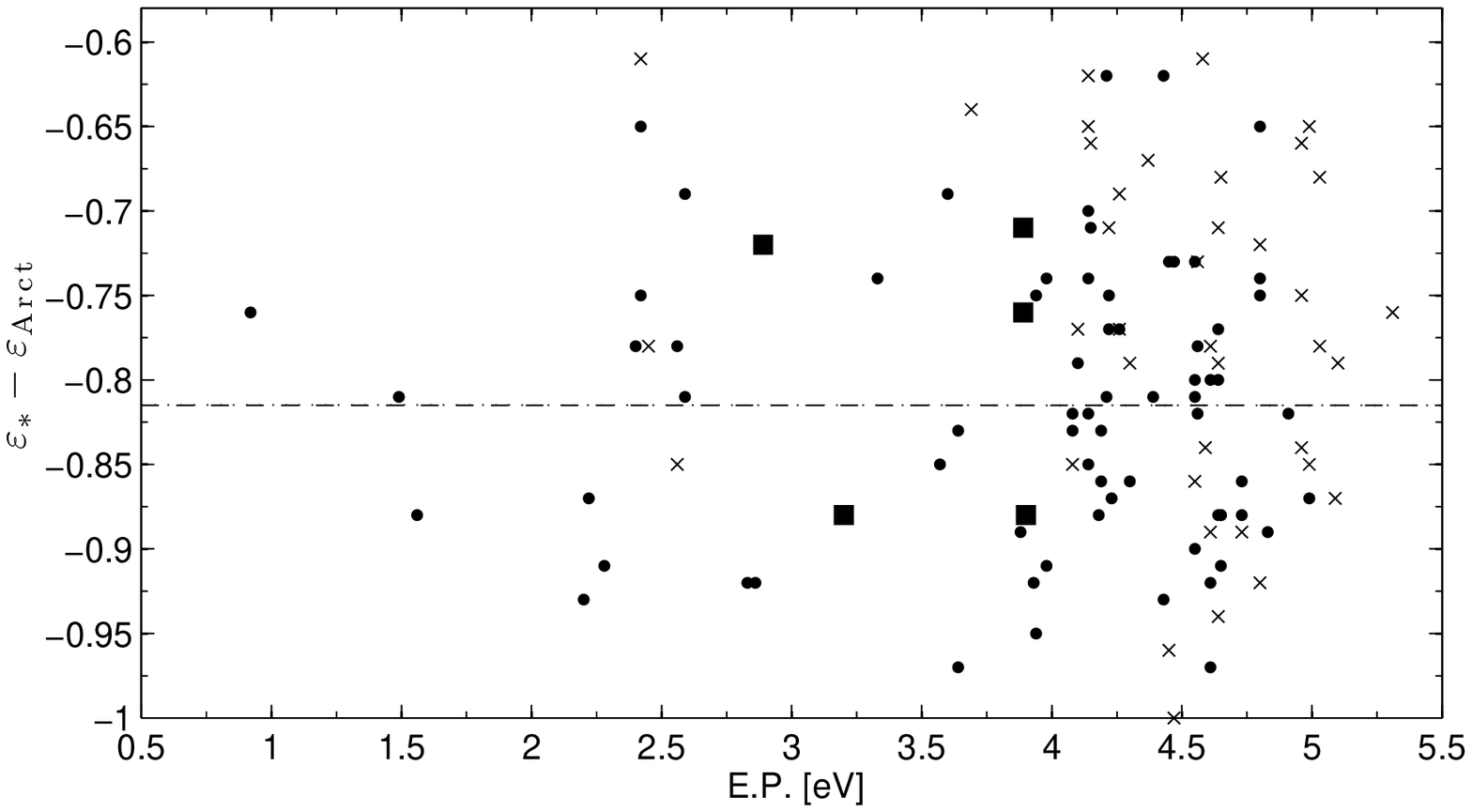}
\includegraphics[angle=0,width=1\hsize]{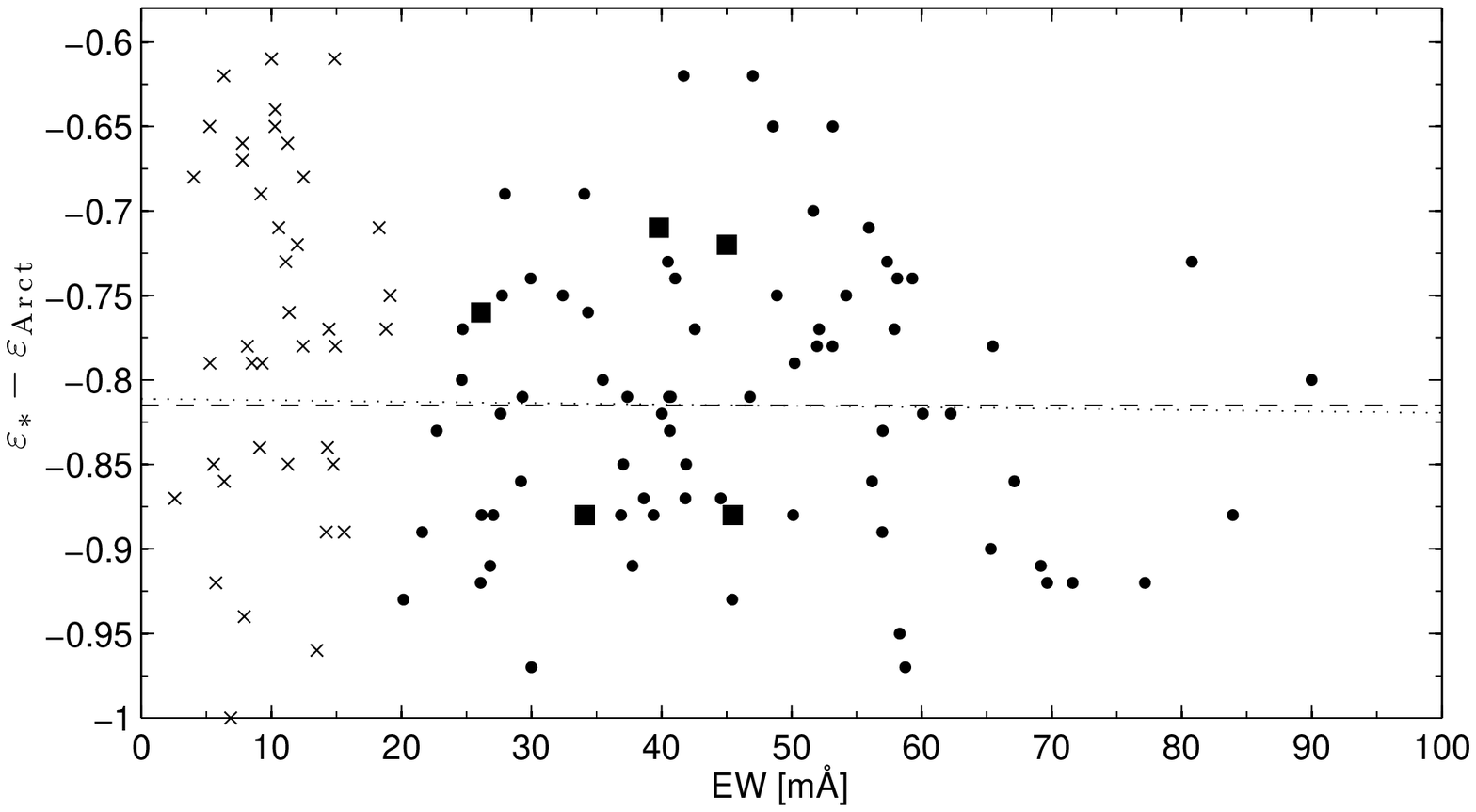}
\end{center}
\caption{EW and EP plot differential to Arcturus, for star M5I-39. 
Small dots (squares) represent \ion{Fe}{1} (\ion{Fe}{2}) lines. Those lines marked as 
crosses have EW$<$20 m\AA\ and were not used in the parameter determinations. Best linear fit and mean abundance are indicated as 
dotted and dashed lines, respectively.}
\end{figure}

Finally, metallicities of the atmospheres [$M$/H] are equated to the [\ion{Fe}{1}/H] abundance 
from the previous iteration step. Table~3 lists the final stellar parameter set that defined our atmospheres 
for the abundance analysis. 
\begin{deluxetable}{cccccc}
\tabletypesize{\scriptsize}
\tablecaption{Atmospheric Parameters}
\tablewidth{0pt}
\tablehead{ \colhead{ID} & \colhead{$T(V-K)$} &  \colhead{$T$(spec)} & \colhead{$T$(average)} & \colhead{log\,$g$}  & \colhead{$\xi$} \\
 & [K] & [K] & [K] & [cm\,s$^{-2}$] & [km\,s$^{-1}$]}
\startdata
M5I-14   &  4278  & 4273 & 4276 & 1.10 & 1.73  \\
M5I-39   &  4315  & 4322 & 4319 & 1.17 & 1.54  \\
M5III-50 &  4386  & 4396 & 4391 & 1.05 & 1.69  \\
M5III-94 &  4207  & 4226 & 4217 & 0.98 & 1.79  \\
M5IV-34  & 4325  & 4289 & 4307 & 1.13 & 1.72  \\
M5IV-82  & 4338  & 4354 & 4346 & 1.24 & 1.82
\enddata
\end{deluxetable}
\subsection{Ionization equilibrium}
As was shown in Paper I,  ionization equilibrium is not established in 
Arcturus, where neutral and ionized species differ by
 [\ion{Fe}{1}/\ion{Fe}{2}] =$-$0.08 dex. 
Recall that the {\em differential} [\ion{Fe}{1}/\ion{Fe}{2}] abundance ratio in the 47 
Tuc giants was found to be +0.08 dex and the results of Paper I indicated that, 
while ionization equilibrium in the cluster stars relative to the Sun could be achieved, 
an equilibrium differential to Arcturus is not satisfied. 

It is noteworthy that we are 
faced with a similar situation in our M5 stars, though in the reversed sense: 
we find an average  difference  [\ion{Fe}{1}/\ion{Fe}{2}] of  $-0.12\pm0.02$ dex for our stars, differential to Arcturus;
this is similar to the difference between neutral and ionized titanium, where
[Ti~I/Ti~II]=$-$0.11$\pm$0.02 dex.  These values are also 
comparable to the abundance difference 
of [Fe~I/Fe~II]=$-$0.09$\pm$0.01 found by Ivans et al. (2001). 
Note that this sense of the non-equilibrium is opposite to that observed in 47 Tuc: the 
abundance of neutral species with respect to the ionized stage is  lower in M5, 
while in 47 Tuc, the neutral lines yielded higher abundances. 
In the mean, ionization equilibrium could be enforced by lowering log $g$ by 0.25  dex on average. 
As was elaborated in Paper I, resorting to systematic errors in mass, luminosity and/or T$_{\rm eff}$ 
can in general explain the apparent over-ionization of iron in our M5 stars relative to Arcturus, although 
the required changes are mostly unphysical. 

If stellar mass was the source of the discrepancy in surface gravity, our targets would need to 
have a mass of 0.46 M$_{\sun}$ on average, which would require an extraordinary large mass-loss
for these first ascent red giants; this seems an undesirable option. 
If changes in log $g$ were due to luminosity effects, the distance modulus of M5 would have 
to be larger by 0.38 mag, i.e., the GC would be more distant by at least 1.6 kpc,  which 
is an unrealistic 
scenario, considering that this value lies 4$\sigma$ above the mean, and 
0.3 mag fainter  than the largest value found in the literature (Gratton et al. 1997). 
Likewise, the photometric accuracy given in Sandquist \& Bolte (2004) is typically  less than a 
few  hundredth of a  magnitude. 
Moreover, the combined contributions of BC uncertainties (at $\sim$$0.05$ mag) and the published 
errors on the reddening of typically 0.01 mag amount to an overall luminosity-effect on log $g$ of
no more than 0.06 dex  which is clearly below the 0.25 dex change in gravity required to explain 
the observed lack of ionization equilibrium of iron. 

Since our color- and excitation temperatures, on our Arcturus scale, 
are in such excellent agreement, it seems 
inadequate to invoke changes in T$_{\rm eff}$ as the source for the non-equilibrium. 
As it turns out, an increase in the temperatures of $\sim$80 K on average would we able to 
resolve the ionization imbalance in our stars (Table~5).  This would, however, result in [Fe/H] 
abundances higher by 0.06$\pm$0.01 dex on average, which is larger than our typical 
random and systematic uncertainties (see Sect.~4.1). 
Moreover, an 80 K discrepancy cannot be accounted for by an increase in the reddening, as such a difference would require 
an E(B$-$V) of the order of 0.07 mag, which clearly provides a poor fit to the CMD and is twice as large 
as the values found in the literature. On the contrary, with a 9.5--10 Gyr isochrone, the best fit 
to the CMD  isochrone  is achieved by lowering the reddening to 0.01, which would rather reflect a 
lower temperature, thereby aggravating the ionization imbalance. 
These arguments confirm that 
temperature effects can be ruled out as the main driver of the non-equilibrium. 

As another possible explanation we explored the sensitivity of the [\ion{Fe}{1}/\ion{Fe}{2}] ratio 
to the adopted $\alpha$-enhancement in the atmospheres. Switching from the opacity distributions 
with an [$\alpha$/Fe] abundance ratio of +0.4 (AODFNEW) to the scaled solar composition (ODFNEW)
is well able to re-establish the ionization equilibrium in the cluster stars on average (see also Table~5) 
and we do in fact find that [\ion{Fe}{1}/\ion{Fe}{2}]$_{\rm ODF}=0.01\pm0.02$ dex. 
Individual giants, however depart from this new equilibrium, and the maximum deviations occur for  
M5I-14 at $-0.07$ dex, and for M5III-94, in which we are now faced with an over-ionization of 0.09 dex.  
As we demonstrate in Section~4.2 (and Table~5), the average enhancement of the M5 stars in the 
$\alpha$-elements (O, Mg, Si, Ca, Ti) amounts to 0.32 dex, and 0.33 dex, if 
the light elements Na and Al are also included. 

Following our abundance analysis we were in a position to 
interpolate the Teramo isochrones to the measured $\epsilon$(O/Fe) and metal
mass fraction, Z, as performed for 47~Tuc in Paper~I.  These,
more appropriate, isochrones should enable an improved interpretation of the photometric
parameters of our M5 stars.  Obviously, it is desirable that photometric and spectroscopic
properties are consistent.
However, the large range of
O/Fe ratios in our sample resulted in ambiguity in the choice of $\epsilon$(O/Fe).
We chose to employ the star with the largest O/Fe ratio, M5~IV-82, which we assumed
reflects the original cluster composition; for this star we computed Z=0.00147 and
$\epsilon$(O/Fe)=1.79 dex.  With Teramo isochrones interpolated to these parameters
we found that only the 14 Gyr isochrone was consistent with the photometry and distance
modulus, to within the 1$\sigma$ uncertainties.  While a reduction in the oxygen
abundance would improve matters slightly, the main source of uncertainty appears to be
the distance modulus.  For an age of 10 Gyr the distance modulus needs to be increased
by 0.23 mag, while a $\sim$0.1 mag increase is required for the 14 Gyr isochrone.  
The 14 Gyr interpolated isochrone indicates an RGB mass of 0.75 M$_{\odot}$ for star
M5~IV-82, which in addition to a 0.10 mag increase in distance modulus decreases
the computed log\,$g$ by 0.09 dex.  If a 10 Gyr isochrone is selected, with a distance
modulus offset of 0.23 mag, and an RGB mass of 0.82 M$_{\odot}$, the log\,$g$ values are
lowered by 0.10 dex.  The slightly lower gravities would reduce the Fe~II
abundances by 0.05 dex, while the Ti~II abundances would be lowered by 0.04 dex;
these shifts should be applied to the Fe~II and Ti~II abundances listed in Table~4.
The gravity corrections result in $\Delta$(FeI$-$FeII)=$-$0.07 dex, from the previous 
$-$0.12 $\pm$0.02 dex, while  $\Delta$(TiI$-$TiII)=$-$0.07 dex, down from 
$-$0.11$\pm$0.03 dex.  Although the ionization equilibrium is improved, the
$\epsilon$(O/Fe) ratio and Z values remain unchanged.  
The above shifts in distance modulus correspond to changes by 1$\sigma$ (for 14 Gyr) and 2.3$\sigma$ (for 10 Gyr) with respect to the 
fiducial value (Section 3.2) and we conclude that   
we are still left with an unresolved ionization imbalance of at least 0.07$\pm$0.02 dex.

Differential non-LTE abundance effects between Arcturus and M5 giants is a potential
source for the observed ionization imbalance.  Qualitatively, we expect the more
metal-poor, more luminous, M5 stars to suffer greater non-LTE over-ionization of neutrals
than Arcturus because the M5 giant atmospheres are more transparent to ionizing UV
radiation and they experience fewer atomic collisions.  The expected sense is for negative
[Fe~I/Fe~II], as observed here.  We note that this is in the opposite sense to 47 Tuc
stars of paper~I, which had a positive [Fe~I/Fe~II] that could not be attributed to
non-LTE over-ionization.  The imbalance, at $\sim$0.07 dex seems not unreasonable, 
although it is larger than would be expected from the ionization plots of 
Fulbright et al. (2007).  On the other hand, since the ionization potential of Fe~I
is significantly higher than for Ti~I one would expect the non-LTE over-ionization
correction for Ti to exceed that of Fe.  

It is conceivable that the lack of ionization equilibrium can be explained by an anomalous 
 He content of the M5 giants: Recall that one (though unlikely) reason for the departure from 
this equilibrium in Arcturus was hypothesized to be an 
unusually high He mass fraction ($Y$=0.40) relative to the standard mixture in the 47~Tuc 
giants. Although we refrain from any quantitative arguments, we briefly note that 
the opposite sense of the ionization imbalance in M5 relative to the 47~Tuc stars  
purports that M5 is rather similar to Arcturus in that this cluster may also exhibit an elevated Y. 
Sandquist \& Bolte (2004) find that the number ratio of AGB to HB stars, $R_2$, which reflects the helium abundance in the convection layers of the stars,  
is higher than found in any other GC (at 0.176) and than predicted by stellar evolution theories  (e.g., Renzini \& Fusi Pecci 1988; Cassisi et al. 2003). 
On the other hand,  the number ratio of HB to RGB stars, R (where a high R signifies a
high initial He content), has been reported to be typical of GCs at the same metallicity
(Sandquist 2000; Salaris et al. 2004).  Moreover, Sandquist (2000) find a canonical Y of
order of 0.2 for M5 from various indicators.  The M5 HB 
stars are on average slightly red and completely unremarkable in HB mean color 
(see Harris 1996), which
is not consistent with a high He content.  An investigation of M5's actual He
enhancements is clearly beyond the scope of the present work.  
\section{Differential abundance results}
Table~4 lists the final abundance ratios determined from the atmospheres under the parameters 
derived in the previous section. The ratios are given relative to \ion{Fe}{1}, except for  [\ion{O}{1}]
and Ti~II, 
which,  due to a similar sensitivity to surface gravity, we state with respect to \ion{Fe}{2}. 
\begin{deluxetable*}{rrrrrrrrrrrr}
\tabletypesize{\scriptsize}
\tablecaption{Abundance Results}
\tablewidth{0pt}
\tablehead {\colhead{} & \multicolumn{3}{c}{M5I-14} & & \multicolumn{3}{c}{M5I-39} & & \multicolumn{3}{c}{M5III-50} \\
\cline{2-4}\cline{6-8}\cline{10-12}
 \raisebox{1.5ex}[-1.5ex]{Ion} & \colhead{{[}X/Fe{]}} & \colhead{$\sigma$}& \colhead{N} & & \colhead{{[}X/Fe{]}} & \colhead{$\sigma$}& \colhead{N} & 
              & \colhead{{[}X/Fe{]}} & \colhead{$\sigma$}& \colhead{N}}
\startdata
{[}Fe\,I/H{]}  &  $-$1.35 & 0.12 & 76 & & $-$1.31 & 0.08 & 72 & & $-$1.41 & 0.10 & 66 \\
{[}Fe\,II/H{]} &  $-$1.16 & 0.09 &   5 & & $-$1.20 & 0.08 &    5 & & $-$1.26 & 0.05 &   4 \\
{[}O\,I{]}\rlap{\tablenotemark{a}}     & 0.15 & \nodata & 1 & & 0.17  & \nodata  & 1 & & 0.11 & \nodata & 1 \\
Na\,I          &  0.22 & \nodata & 1 & & 0.28 & 0.04  & 2  & & 0.28 & 0.01 & 2  \\
Mg\,I          &  0.40 & 0.07  & 7 & & 0.36 & 0.07 & 5 & & 0.30  & 0.06 & 4 \\
Al\,I          & 0.62  & 0.07 & 4 & & 0.56  & 0.08  & 4 & & 0.77 & 0.02 & 4 \\
Si\,I          & 0.38 & 0.10 & 5 & & 0.33 & 0.01  & 5 & & 0.40 & 0.09  & 5 \\
Ca\,I          &  0.35 & 0.08 & 8 &  & 0.50 & 0.10 & 8 &  & 0.33  & 0.12 & 9 \\
Ti\,I          & 0.22 & 0.11 & 7 & & 0.28 & 0.09 & 9 & & 0.05 & 0.04 & 3\\
Ti\,II\rlap{\tablenotemark{a}}          &  0.09 & 0.19 & 3 & & 0.33 & 0.08 & 3 & & 0.11  & 0.00 & 2 \\
\hline
\colhead{} & \multicolumn{3}{c}{M5III-94} & &\multicolumn{3}{c}{M5IV-34} & & \multicolumn{3}{c}{M5IV-82} \\
\cline{2-4}\cline{6-8}\cline{10-12}
 \raisebox{1.5ex}[-1.5ex]{Ion} & {[}X/Fe{]} & $\sigma$ & N  & & {[}X/Fe{]} & $\sigma$ & N & & {[}X/Fe{]} & $\sigma$ & N   \\
\hline					      
{[}Fe\,I/H{]}   & $-$1.23 & 0.13 & 81 & & $-$1.31 & 0.09 & 74 & & $-$1.37 & 0.11 & 72 \\
{[}Fe\,II/H{]}  & $-$1.15 & 0.06 &   4 & & $-$1.25 & 0.03 &   4 &  & $-$1.24 & 0.04 & 5 \\
{[}O\,I{]}\rlap{\tablenotemark{a}}       & $-$0.03 & \nodata & 1 & & 0.19 & \nodata & 1 & & 0.59  & \nodata & 1 \\
Na\,I           &  0.34 & 0.08 & 2 & & 0.11  & 0.02 & 2 & & $-$0.11  & 0.03 & 2  \\
Mg\,I           & 0.48 & 0.09 & 6 & & 0.48 & 0.05 & 5 & & 0.40 & 0.09 & 5 \\
Al\,I           & 0.71  & 0.07 & 4 & & 0.47 & 0.06 & 4 & & 0.25 & 0.09 & 4 \\
Si\,I           & 0.48 & 0.16 & 5 & & 0.35 & 0.10 & 5 & & 0.38 & 0.08 & 5 \\
Ca\,I           & 0.44  & 0.08  & 8 & & 0.33  & 0.07  & 8 & & 0.37 & 0.07 & 8 \\
Ti\,I          & 0.26 & 0.08 & 8 & &  0.27 & 0.08 & 9 & &0.28  & 0.15 & 7 \\
Ti\,II\rlap{\tablenotemark{a}}          &  0.25 & 0.17 & 4 & & 0.28 & 0.06 & 4 & & 0.21 & 0.07 & 4
\enddata
\tablenotetext{a}{Relative to \ion{Fe}{2}.}
\end{deluxetable*}
\subsection{Abundance errors}
In order to quantify the systematic errors on the elemental abundances that originate in 
uncertainties in the stellar atmosphere parameters, we performed an identical 
standard error analysis as in Paper I. 
For this purpose, we varied  the atmosphere parameters by the following conservative uncertainties 
and re-computed new abundances: (T$\pm$50K; log $g \pm$ 0.2 dex; $\xi \pm$ 
0.1 km\,s$^{-1}$; [$M$/H]$\pm$0.1 dex). The effect on the final abundance ratios is displayed 
in Table~5, for two stars (M5III-94 and M5IV-82) with parameters that cover a large range
in temperature and gravity.
Also, their spectra span the full representative range in S/N ratios 
used in this study (Tables~1,3). 
In addition, we computed atmospheres with Solar scaled opacity distributions, ODFNEW, 
which reduces the $\alpha$-enhancement in the input models by 0.4 dex. The resulting abundance 
variations are found in Table~5 in the column labeled ``ODF''. 
\begin{deluxetable*}{lrrrrrrrrrr}
\tabletypesize{\scriptsize}
\tablecaption{Error analysis for the giants M5III-94 and M5IV-82.}
\tablewidth{0pt}
\tablehead{
\colhead{} & \colhead{} & \multicolumn{2}{c}{$\Delta$T$_{\rm eff}$} & \multicolumn{2}{c}{$\Delta\,\log\,g$} & \multicolumn{2}{c}{$\Delta\xi$} 
& \multicolumn{2}{c}{$\Delta$[M/H]} & \colhead{} \\
 &  \raisebox{1.5ex}[-1.5ex]{Ion}  & \colhead{$-$50\,K}  & \colhead{+50\,K} & \colhead{$-$0.2\,dex} & \colhead{+0.2\,dex} & \colhead{$-$0.1\,km\,s$^{-1}$} & \colhead{+0.1\,km\,s$^{-1}$} & \colhead{$-$0.1\,dex} & \colhead{+0.1\,dex} &  \raisebox{1.5ex}[-1.5ex]{ODF} 
 }
\startdata
& Fe\,I    & $-$0.02 &    0.03 & $-$0.01 &    0.04 &    0.02 & $-$0.02 & $-$0.01 &    0.01 & $-$0.03 \\
& Fe\,II   &    0.06 & $-$0.04 & $-$0.07 &    0.13 &    0.02 & $-$0.03 & $-$0.04 &    0.03 & $-$0.13 \\
&{[}O\,I{]}& $-$0.03 &    0.02 & $-$0.08 &    0.09 & $<$0.01 & $-$0.01 & $-$0.04 &    0.03 & $-$0.12 \\
& Na\,I    & $-$0.04 &    0.04 &    0.01 & $-$0.01 &    0.01 & $-$0.01 & $<$0.01 & $-$0.01 &    0.02 \\
& Mg\,I    & $-$0.01 &    0.02 &    0.01 &    0.02 &    0.02 & $-$0.02 & $<$0.01 & $<$0.01 & $-$0.01 \\
 \raisebox{1.5ex}[-1.5ex]{M5III-94} & Al\,I    & $-$0.03 &    0.03 &    0.01 & $<$0.01 &    0.02 & $-$0.01 &    0.01 & $<$0.01 &    0.01 \\
& Si\,I    &      0.04  &  $-$0.02 &  $-$0.02 &  0.07 &   0.02 &  $-$0.02 &  $-$0.02 &   0.02  & $-$0.06\\
& Ca\,I    &    $-$0.06 &   0.06 &    0.02 &   $<$0.01 &   0.06 &  $-$0.06 &   0.01 &  $-$0.01 &  $<$0.01\\
& Ti\,I    &     $-$0.10 &  0.10 & $<$0.01 &  $-$0.01 &   0.03 &  $-$0.02 &   0.01 & $<$0.01& $-$0.01 \\
& Ti\,II   &      0.02 &  $<$0.01  &  $-$0.06  &  0.10  &  0.04  & $-$0.04  & $-$0.03  &  0.03   & $-$0.11 \\
\hline
& Fe\,I    & $-$0.03 &    0.03 & $-$0.01 &    0.04 &    0.02 & $-$0.01 &    0.01 & $<$0.01 & $-$0.01 \\
& Fe\,II   &    0.03 & $-$0.02 & $-$0.11 &    0.08 &    0.02 & $-$0.02 &    0.04 & $-$0.04 & $-$0.11 \\
&{[}O\,I{]}& $-$0.03 &    0.02 & $-$0.09 &    0.07 & $<$0.01 & $-$0.01 & $-$0.04 &    0.03 & $-$0.12 \\
& Na\,I    & $-$0.04 &    0.04 &    0.02 & $<$0.01 &    0.01 & $-$0.01 &    0.01 & $<$0.01 &    0.02 \\
& Mg\,I    & $-$0.03 &    0.02 & $<$0.01 & $-$0.01 &    0.02 & $-$0.02 & $<$0.01 & $<$0.01 & $-$0.01 \\
 \raisebox{1.5ex}[-1.5ex]{M5IV-82} & Al\,I    & $-$0.03 &    0.03 &    0.01 & $-$0.01 & $<$0.01 & $-$0.01 &    0.01 & $-$0.01 &    0.01 \\
& Si\,I    & 0.01 & $-$0.01 & $-$0.05 & 0.03 & 0.01 & $-$0.01 & $-$0.02 & 0.02 & $-$0.05 \\
& Ca\,I    & $-$0.06 & 0.05 & $<$0.01 & $-$0.02 & 0.05 & $-$0.04 & 0.01 &$<$0.01  & 0.01  \\
& Ti\,I    &  $-$0.08 &   0.08 &   0.02 &  $<$0.01 &   0.01  & $-$0.01 &   0.01 &  $-$0.01 &   0.02 \\
& Ti\,II   &  $<$0.01 & $<$0.01 & $-$0.09 &  0.06 &  0.03 &  $-$0.03 &  $-$0.04 &   0.03 &  $-$0.10
\enddata
\end{deluxetable*}

In practice, the r.m.s. scatter of 28 K in the comparison of  T(V$-$K) versus excitation temperature 
indicates a 1$\sigma$ random error on either indicator of $\sim$23 K. 
If we glean the systematic uncertainty of 30 K for Arcturus from Paper I, and add 
these values  in quadrature we obtain a total error on T$_{\rm eff}$ of 35 K. 

By accounting for errors on  distance modulus, reddening, V-magnitude, 
BC and stellar mass (Sect.~3.2) we estimate a surface gravity uncertainty of 0.06 dex. 
Moreover, we assume a $\Delta\xi$ of 0.05 km\,s$^{-1}$, based on the slope uncertainty 
of the EW vs. $\varepsilon$(\ion{Fe}{1}) plots, and a 0.05 error on the models'  metallicity [$M$/H]. 
Finally, we adopted an error of 0.1 dex on [$\alpha$/Fe], which  
corresponds to 1/4 of the difference when ODFNEW is used as opposed to the AODFNEW atmospheres, 
and which is of the order of a typical 1$\sigma$ scatter in the derived $\varepsilon(\alpha)$ 
abundances (Table~4). 
Interpolating from Table~5, all these contributions are finally added in quadrature to obtain 
an {\em upper limit} for the total uncorrelated systematic 
abundance uncertainty. Note, however, that the actual errors on our abundance ratios 
are probably smaller, due to the covariances of all atmospheric parameters (McWilliam et al. 1995b).  
As a result, the total 1$\sigma$ systematic errors on the [\ion{Fe}{1}/H] and [\ion{Fe}{2}/H] 
abundance ratios are 0.03 and 0.06 dex, respectively, and typically 0.02--0.06 dex for the 
$\alpha$-elements. 

Table~4 additionally lists the number of features $N$  that were measured to derive the elemental 
abundances, and the statistical error in terms of the 1$\sigma$ scatter of our measurements from 
individual lines. 
This error component is negligible compared to those from the atmospheric uncertainties 
for \ion{Fe}{1}, where a sufficient number of lines is detectable. For the other chemical elements with only a handful of measurable transitions, however, this 
statistical scatter will dominate the error budget. We find random r.m.s. scatters ranging from 0.01--0.20 
dex, corresponding to a mean error of 0.04 dex per line. 
\subsection{Abundance ratios}
\subsubsection{Iron}
From our six targets   we derive a mean value of [\ion{Fe}{1}/H]=
$-1.33\pm0.03\pm0.03$ dex (random and systematic error). The former, 
statistical, error is simply the standard deviation of the mean. At 0.06 dex, the 
intrinsic star-to-star scatter within M5 is 
slightly larger 
than the measurement errors estimated  above. 
The cluster mean from this analysis 
agrees well with the calcium triplet based value from low resolution studies 
(Zinn \& West 1984; Koch et al. 2006) and with high-resolution measurements found 
in the literature. For instance, the  study of 36 RGB and AGB stars in M5 
by Ivans et al. (2001) finds a mean of $-1.34$ with a scatter of 0.06 dex, although this value 
appears to depend on evolutionary status, where the AGB sample exhibits a mean of 
$-1.44$ and the red giants have a higher mean of $-1.29$ dex.   Also our one AGB candidate, M5III-50, 
has the lowest iron abundance of our sample, in accord with the finding of Ivans et al. (2001). 
Removing this star from the statistics has only a marginal influence and the mean [\ion{Fe}{1}/H] 
of the pure RGB sample remains at $-1.31\pm0.02$ dex. 
Likewise, Carretta et al. (2009a) find a mean of $-$1.35 dex and only a low r.m.s. scatter of 0.02 dex from a comprehensive analysis of 136 stars. 

In Fig.~3 we illustrate the  deviations of our values from  the literature.  
These differences are also tabulated in Table~6.
\begin{figure}
\begin{center}
\includegraphics[angle=0,width=1\hsize]{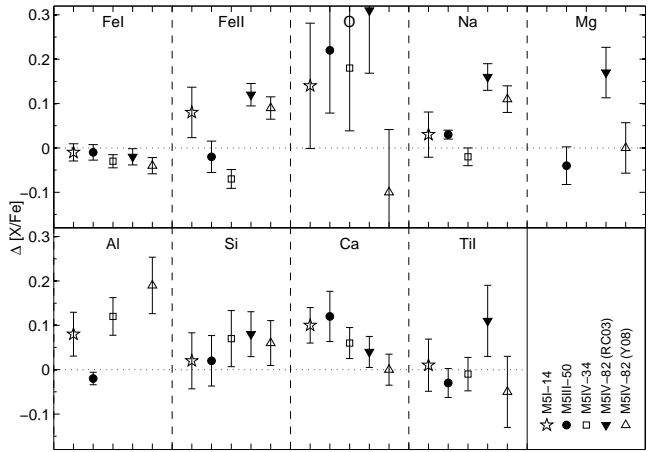}
\end{center}
\caption{Deviations of our abundance scale from literature data in the sense $\Delta = X_{\rm This~work} - X_{\rm  Lit.}$. The comparison includes 
overlapping stars from Ivans et al. (2001); Ramirez \& Cohen (2003; RC03); Yong et al. (2008a,b; Y08).}
\end{figure}
Those stars, for which absolute (i.e., non-differential) measurements 
are available in the literature (see references in Sect.~1), agree with our data on average and we find a mean difference of $-0.02\pm0.01$ dex for \ion{Fe}{1}. 
This is a good confirmation of our new abundance scale. 
Such an agreement of the abundance scales is different from our findings for 47~Tuc in Paper~I, where  the newly established differential scale differs 
from previous high-resolution studies that relied on atomic parameters by $\sim$0.1 dex on average. 
\begin{deluxetable*}{rccccccc}
\tabletypesize{\scriptsize}
\tablecaption{Deviations of [X/Fe] from literature data (Fig~3).}
\tablewidth{0pt}
\tablehead { 
\colhead{} & \colhead{M5I-14}  &  \colhead{M5III-50} &  \colhead{M5IV-34} & \multicolumn{2}{c}{M5IV-82} & \colhead{} & \colhead{} \\
\cline{2-4}\cline{5-6}
\raisebox{1.5ex}[-1.5ex]{Ion} & \multicolumn{3}{c}{Ivans et al. (2001)} & 
\colhead{(RC03)} & \colhead{(Y08)} & \raisebox{1.5ex}[-1.5ex]{Mean}  & \raisebox{1.5ex}[-1.5ex]{$\sigma$}  }
\startdata
{[}Fe\,I/H{]}  &      $-$0.01 &  $-$0.01 &     $-$0.03 & $-$0.02    &  $-$0.04 &  $-$0.02 & 0.01 \\
{[}Fe\,II/H{]} &     \phs0.08 &  $-$0.02 &     $-$0.07 & \phs0.12    & \phs0.09 & \phs0.04 & 0.08 \\ 
{[}O\,I{]}     &     \phs0.14 & \phs0.22 &     \phs0.18 &  \phs0.31    &  $-$0.10 & \phs0.15 & 0.15 \\
Na\,I          &     \phs0.03 & \phs0.03 &     $-$0.02 & \phs0.16    & \phs0.11 & \phs0.06 & 0.07 \\
Mg\,I          &  \phs\nodata &  $-$0.04 & \phs\nodata & \phs0.17    &  \phs0.00 & \phs0.04 & 0.11 \\ 
Al\,I          &     \phs0.08 &  $-$0.02 &    \phs0.12 & \phs\nodata & \phs0.19 & \phs0.09 & 0.09 \\
Si\,I          &     \phs0.02 & \phs0.02 &    \phs0.07 & \phs0.08    & \phs0.06 & \phs0.05 & 0.03 \\
Ca\,I          &     \phs0.10 & \phs0.12 &    \phs0.06 & \phs0.04    & \phs0.00 & \phs0.06 & 0.05 \\
Ti\,I          &      \phs0.01 &  $-$0.03 &     $-$0.01 & \phs0.11    &  $-$0.05 &  \phs0.01 & 0.06
\enddata
\end{deluxetable*}

The  discrepancies in the \ion{Fe}{2} values are only marginal, at  0.04 dex, although there is a notably large 1$\sigma$ scatter of 0.08 dex. Yet this 
should not cause any concern: 
The studies of Yong et al. (2008a) forced ionization equilibrium by adjusting log $g$ to yield 
consistent \ion{Fe}{1} and II abundances, while our study and those of Ivans et al. (2001)   
and  Ramirez \& Cohen (2003) rely on the photometric gravities and take the \ion{Fe}{1} to \ion{Fe}{2} discrepancy at face value. 
If we removed  the Yong et al. measurement from the statistics, there would still be a marginal  deviation of 0.03 dex (1$\sigma$-scatter of 
0.09 dex) so that the adopted gravity scales are not a likely explanation of the scatter.
We also note  that Ivans et al. (2001) used MARCS models with 
scaled solar composition, whilst employing a higher model [Fe/H] in order to 
compensate for the effect of  the  expected enhanced [$\alpha$/Fe] ratios (Fulbright \& Kraft 1999). 
The methodology difference between the model atmospheres used by Ivans et al. (2001) and this work could 
reasonably affect the electron number densities employed in the abundance calculations, which would result in 
systematically different [Fe I/Fe II] ratios. 
\subsubsection{Alpha-elements --- O, Mg, Si, Ca, Ti}
Fig.~4 illustrates the derived abundance ratios in a statistical box plot, which shows the mean and interquartile ranges for each element. 
The respective abundances are based on typically 5--8 sufficiently strong lines. 
Much information about the detailed abundance distribution of M5 is found in the literature and we do not opt to 
repeat the main arguments from those sources (see Ivans et al. 2001; Ram\'irez \& Cohen 2003; Yong et al. 2008a,b), which  
confirmed  M5 as a typical Galactic halo cluster with an indication of star-to-star scatter in a few (light) elements. 
Instead, we  rather  focus on the comparison of our new abundance scale to those data  
and discuss a few interesting individual cases. 
\begin{figure}
\begin{center}
\includegraphics[angle=0,width=1\hsize]{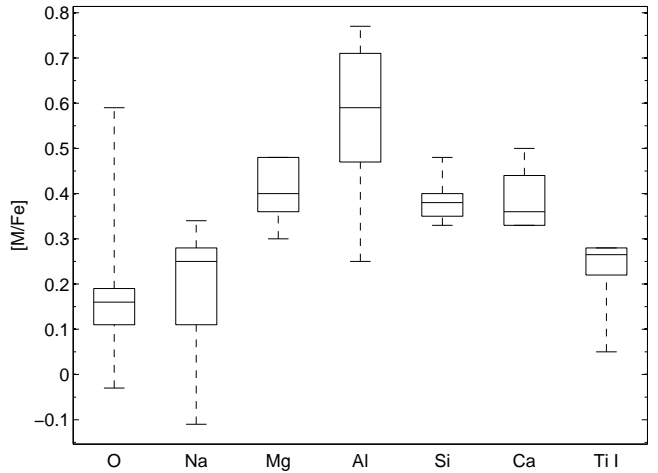}
\end{center}
\caption{Boxplot of the differential abundance ratios in our M5 sample.}
\end{figure}

Despite a reasonable agreement in each element for a couple of the stars that overlap with the literature, some of the elements in this comparison 
are discordant by more than  expected from the measurement uncertainties. 
This is unlikely to be caused by differences in the adopted gravities, since both Ivans et al. (2001) and Ram\'irez \& Cohen (2003) used a very similar treatment of ionization equilibrium as we did and employed very similar values of the surface gravities. 
It is, rather, very likely that the reason for these discrepancies lies in the different nature of the studies, i.e., the use of log$gf$ values versus our differential method, and differences in the line lists. 
Overall, the mean values are in fair agreement; 
in particular we find 	that our differential [$\alpha$/Fe] element ratios are higher on average than the five literature 
values in common by  0.04$\pm$0.06 dex (Mg), 0.05$\pm$0.01 dex (Si), 0.06$\pm$0.02 dex (Ca), and 0.01$\pm$0.03 dex (Ti), respectively. 

M5 is a typical halo cluster in that it is enhanced in the $\alpha$ elements Mg, Si, and Ca to the plateau halo value of +0.4 dex (note that 
Ti is less enhanced by ca. 0.1 dex), while 
the [O/Fe] abundances are lower, with an average [O/Fe] of 0.20$\pm$0.09 dex. This is the opposite trend seen in the more metal rich GC 47~Tuc 
(Paper~I), which shows oxygen abundances higher by 0.2 dex compared to the other $\alpha$-element enhancements, and which was argued 
to be consistent with 47~Tuc's properties similar to the Galactic bulge.  

The $\alpha$-elements are believed to 
be over-produced, relative to Fe, in supernovae (SNe) of type II, which relate to the death of
massive, therefore short-lived, stars. While Mg and O are  formed in the hydrostatic nuclear 
burning in the SNe II progenitors only, Si, Ca, and Ti are rather assigned to the 
explosive nucleosynthetic phase of the SNe II (e.g., Woosley 
\& Weaver 1995).\footnote{Small, but not negligible, amounts of these explosive alphas are also 
predicted to be produced in type~I SNe (e.g. Nomoto et al. 1984, Iwamoto et al. 1999).}  
Thus, differences in abundance trends of hydrostatic versus explosive alpha-elements can be
expected.  Alpha-element trend differences have been seen for [Mg/Fe]
compared with [Si,Ca,Ti/Fe] versus [Fe/H] by Fulbright et al. (2007) and paper~I.

Therefore, we plot in 
Fig.~5 (bottom panels) the difference of Ca and Si, Ca and Ti~I, respectively, 
versus metallicity.
Those plots and the following figures illustrate our M~5 data in comparison
with the Galactic  
 bulge (Fulbright et al. 2007) and halo stars (Nissen \& Schuster 1997; 
Hanson et al. 1998; Fulbright 2000; Stephens \& Boesgaard 2002; Fulbright \& Johnson 2003). 
\begin{figure}
\begin{center}
\includegraphics[angle=0,width=1\hsize]{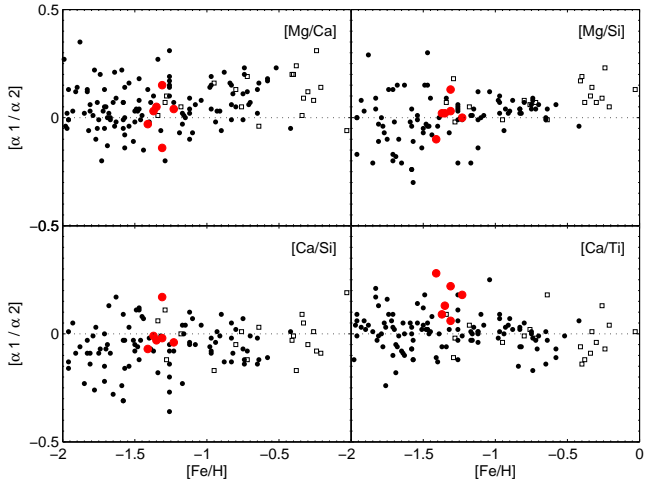}
\end{center}
\caption{Various $\alpha$-to-$\alpha$-element ratios for M5 (red filled circles),  Galactic bulge (open squares), and 
halo (points) stars. See text for references.}
\end{figure}

The [Ca/Si] ratio is fully consistent with zero, as found in all Galactic components at these moderately low 
metallicities, with only little scatter (r.m.s. of 0.08 dex). This confirms that the explosive $\alpha$-elements indeed 
trace each others' element trends. 
One star, M5I-39, shows a [Ca/Si] ratio that is higher by almost 0.2 dex than the zero-average. This star also shows the lowest [Mg/Ca] ratio, 
while its [Ca/Ti] and [Mg/Si] ratios are fully compatible with the sample means. This may point to an unusually high enhancement of this star in Ca alone. 

At a mean of +0.10 dex with a 1$\sigma$-scatter of 0.06 dex, the [Ca/Ti] ratio is
slightly higher than the Galactic values.  It is possible that this is due to 
non-LTE over-ionization of Ti~I.
We note that this $\sim$0.10 dex enhancement in the [Ca/Ti] abundance ratio would be
reduced to $\sim$0.03 dex assuming the presumed Ti~I non-LTE correction of 0.07 dex
required to obtain ionization equilibrium (see section 3.3).  Qualitatively, however, 
the non-LTE correction should be smaller for Fe~I than for Ti~I, due to the higher
ionization potential of Fe~I (7.90eV versus 6.83 eV). That we measure the same
ionization equilibrium deficit for Ti and Fe suggests that the continuum formation is
more likely the source than non-LTE, in which case the $\sim$0.10 dex [Ca/Ti] enhancement
is probably real.
In this regard, it is also interesting to note that 
both the [Ca/Si] and [Ca/Ti] values are higher by about 0.1 dex than found in the metal rich 47~Tuc in Paper~I. 
A glance at the [Mg/Ca] plot in the top panel of Fig.~5 reveals that this ratio is in full agreement with the value of $\sim$0 found in the Galactic halo and bulge, 
except for the slightly discordant star M5I-39. 
As the Mg versus Ca production the SNe II is a delicate function of progenitor mass, this shows that there is no need to invoke any enrichment through very massive  
stars in M5, as found, e.g., in low-mass environments such as the (ultra-) faint dwarf spheroidal galaxies (Koch et al. 2008; Feltzing et al. 2009). 
This holds for the observed [Mg/Si] ratios as well, which are compatible with zero in our M5 stars (at a mean and 1$\sigma$ scatter of 0.02$\pm$0.07 dex). 
\subsubsection{Light elements --- O, Na, Al}
The oxygen abundances are solely based on the [\ion{O}{1}] 6300\AA~line that we carefully 
deblended from telluric absorption. The transition at 6363\AA~is   
generally too weak in our spectra and further affected by a Ca auto-ionization feature. 
For the 6300 [O~I] line we assign an upper limit of the uncertainty on the derived [O/Fe] ratio of 
0.10 dex. The final values for the five stars in common deviate from the literature values 
by 0.15 dex on average, with a 1$\sigma$ scatter of 0.15 dex. 
Most authors performed profile matching, while our values rely on EW measurements. 
Given the good agreement of the gravities between all studies in question, and considering the order of magnitude 
of the change in the O/Fe ratio under log\,$g$ variations (Table~5), it is unlikely that all of the deviation in [O/Fe] 
is due to the adopted gravity scales. 
The oscillator strengths for the O-lines employed in the comparison data all agree to within better than 0.05 dex, which 
leads to differences in [O/Fe] of 0.04 dex at most (although we note that Ram\'irez \& Cohen 2003 used 
in addition the 7771\AA-triplet). Therefore we conclude that the deviation of our abundances from those in the literature are 
a consequence of the fundamentally different, to wit differential, analysis method of this work. 

The Na abundances for our stars were derived from the EWs of the 
6154\AA~doublet only (cf. Ivans et al. 2001). The strong lines at 5682, 5688\AA~are heavily blended 
with other metal lines and in the present, truly differential, EW study we wish to avoid any 
dependence on spectral syntheses that would resolve such blends. 
We find a mean [Na/Fe] of 0.19$\pm$0.07 dex, which agrees well with the literature measurements. 
While our sodium abundances do not deviate by more than 0.03 dex from the values 
of Ivans et al. (2001), our [Na/Fe] for star M5IV-82 is higher by 0.16 (0.11) dex than 
stated by Ram\'irez \& Cohen (2003) and Yong et al. (2008b), respectively (Fig.~3). 
All of these authors have included the strong Na-D lines in their computations so that 
the observed difference in [Na/Fe] might be due to our removal of this feature from our line list. 
On the other hand, this star is also characterized by a [Na/Fe] ratio that is lower 
than the cluster mean by 0.4 dex (2.4$\sigma$). 
This finding is, however, consistent with the high oxygen abundance of this star, which leads to 
a distinctly low [Na/O] ratio, see Fig.~6. 
\begin{figure}
\begin{center}
\includegraphics[angle=0,width=1\hsize]{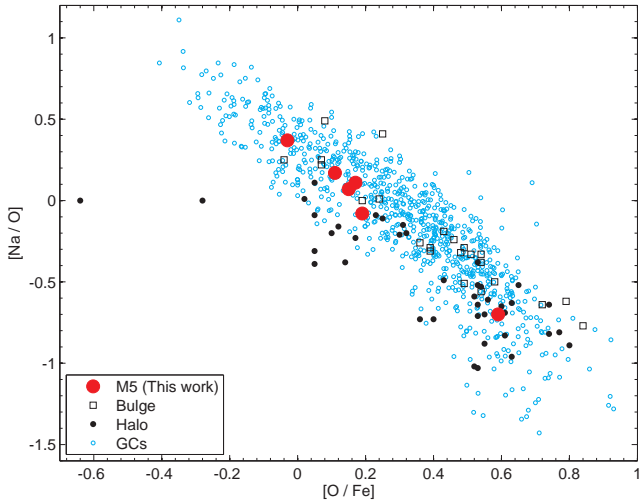}
\end{center}
\caption{Anti-correlation of Na and O abundances for the Galactic bulge and halo (see text for references), Galactic GCs from Carretta et al. (2009b; blue open circles); 
 and the M5 data from this work (red solid circles).}
\end{figure}

In this Figure, we illustrate the well-defined Na-O anti-correlation 
found in GC stars (e.g., Gratton et al. 2004; Carretta et al.  2009b,c),  in comparison with the 
Galactic stars also indicated in Fig.~5. 
While the halo  distribution does not show any Na-O relations,   
M5 follows very closely the trend outlined by the Galactic halo clusters, which is in accord with the finding of Ivans et al. (2001).
In particular, we can assign our star M5IV-82 to the first-generation, ``primordial'' component that shows (low) Na and (high) O abundances similar to the halo field stars and therefore abundance patterns that are consistent with SN nucleosynthesis (Carretta et al. 2009b), while the remainder shows the elevated [Na/Fe] ratios  and lower O-abundances typical of the material processed through proton capture nucleosynthesis in the second generation of the cluster stars. 

Aluminum abundance ratios in our stars were measured from four relatively weak (20--40 m\AA) lines, around 6696 and 7835\AA.
As a result, we see evidence for a broad range in the Al/Fe ratio of about 0.5 dex. 
In Fig.~7 we parallel the Mg/Al correlation with the Na-O anti-correlation discussed above. As before, it is evident that M5IV-82, having the highest Mg/Al ratio in our sample, reflects the original, chemically unprocessed composition of M5. 
Therefore, our {\em differential abundances} confirm the 
occurrence of star-to-star variations in the 
light elements due to the canonical proton-capture synthesis processes seen in other metal poor Galactic halo GCs. 
\begin{figure}
\begin{center}
\includegraphics[angle=0,width=1\hsize]{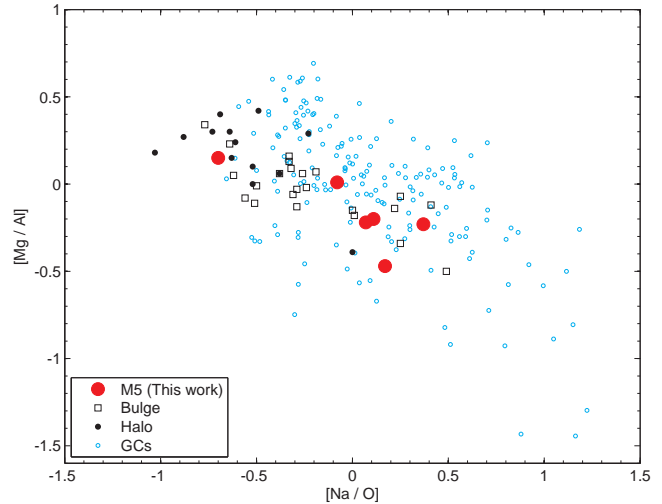}
\end{center}
\caption{Similar to Fig.~6, but for the  correlation of the Mg/Al abundance ratio with the Na/O ratio. GC data (open circles) were taken from Carretta et al. (2009c).}
\end{figure}
\section{Summary \& Discussion}
Here we have determined chemical abundance ratios in five red giants and one AGB star in the moderately metal poor 
Galactic halo globular cluster M~5. 
One major improvement of our measurements of iron, Na and Al, and the $\alpha$-element-to-iron ratios (O, Mg, Si, Ca, Ti/Fe) 
compared to previous works on this well-studied GC is the {\em differential} line-by-line abundance relative to a reference star of similar stellar parameters, to wit, Arcturus.  As a consequence, inevitable uncertainties in the model atmospheres and potentially erroneous atomic data were efficiently reduced, by which we could attain a high degree of accuracy of our abundance results. 
Our derived mean LTE [Fe/H] of $-1.33\pm0.03$$\pm0.03 $ dex is fully consistent with recent results in the literature, while a number of the $\alpha$-element ratios we find are higher by $\sim$0.04 dex on average than literature values owing to the different analysis techniques employed in each of the sources.  
At first glance this may seem to imply that log\,$gf$-based abundances can provide an equally well suited abundances scale. We point out, however, that the discrepancies between the differential and absolute values found for 47~Tuc  (Paper~I)  indicate that such an agreement is not necessarily generally granted so that any such comparison 
of GC abundance scales must be considered on a case-by-case basis. 

Following our accurate new measurements for the metal rich GC 47~Tuc in Paper~I, the present study is the second step towards an accurate differential abundance scale for Galactic GCs, which  will enable future age dating to an unprecedented precision and establish M5 as a template of moderately metal poor clusters. 
The necessity for an accurate age determination is underscored by the use of M5 as a CMD template for characterizing faint, remote, systems for which hardly any 
spectroscopic information is achievable (cf. Koch et al. 2009). In this context,  Stetson et al. (1999) argue that M5 should be younger than the old halo cluster M3 by $\sim$1 Gyr, while Vandenberg (2000) finds M5 ``to be slightly older than most clusters having similar metallicities''. 

While the explosive $\alpha$-elements Si and Ca trace each other, confirming their origin in  
explosive nucleosynthesis, the [Ca/Ti] 
ratio is higher than in the Galactic halo; M5's solar [Mg/Ca] ratio then indicates
that 
the chemical enrichement history of the M5 material is consistent with a normal mix
of SN progenitor masses.
Our data show clear evidence of a Na-O  anti-correlation, as well as  a broad range of  Al abundances, all of which implies the occurrence of proton capture processes in the cluster stars. One star of our sample, however, represents the primordial, unprocessed composition of a first generation of stars. 
All other $\alpha$-elements (Mg, Si, Ca, and Ti) show  only little star-to-star (1$\sigma$) scatter 
 (of the order of 0.05--0.07 dex, compared to 0.16 and 0.21 dex for Na and O, respectively) that is compatible with the measurement errors.  
 This finding is in agreement with Carretta et al. (2009a), who 
report a very low r.m.s. scatter in their iron abundances, and which bolsters that this system has formed and evolved in a chemically homogeneous 
environment. 

Its kinematics indicates that M5 spends 90\% of its  orbit at Galactocentric distances outside of 10 kpc (Dinescu et al. 1999) and it comes as little surprise that,  
all in all, this cluster overlaps with the canonical trends delineated by Galactic halo field and GC stars at the same metallicity so that it can be considered a typical (present-day) inner halo cluster  (cf. Koch et al. 2009). 
\vspace{-0.5cm}
\acknowledgments
We gratefully acknowledge funding for this work from a NASA-SIM key project grant, entitled ``Anchoring 
the Population II Distance Scale: Accurate Ages for Globular Clusters and Field Halo Stars''. 
AK acknowledges support by an STFC postdoctoral fellowship. 
This research has made use of the NASA/ IPAC Infrared Science Archive, which is operated by the Jet
Propulsion Laboratory, California Institute of Technology, under contract with the National Aeronautics
and Space Administration
\end{document}